\documentclass[conference]{IEEEtran}
\IEEEoverridecommandlockouts
\usepackage{cite}
\usepackage{amsmath,amssymb,amsfonts}
\usepackage{algorithmic}
\usepackage{graphicx}
\usepackage{textcomp}
\usepackage{xcolor}
\usepackage{hyperref}

\def\BibTeX{{\rm B\kern-.05em{\sc i\kern-.025em b}\kern-.08em
    T\kern-.1667em\lower.7ex\hbox{E}\kern-.125emX}}
\begin{document}

\title{Requirements are All You Need: From Requirements to Code with LLMs\\
}

\author{\IEEEauthorblockN{Bingyang Wei}
\IEEEauthorblockA{\textit{Department of Computer Science} \\
\textit{Texas Christian University}\\
Fort Worth, USA \\
b.wei@tcu.edu}
}

\maketitle

\begin{abstract}
The pervasive use of textual formats in the documentation of software requirements presents a great opportunity for applying large language models (LLMs) to software engineering tasks. High-quality software requirements not only enhance the manual software development process but also position organizations to fully harness the potential of the emerging LLMs technology. This paper introduces a tailored LLM for automating the generation of code snippets from well-structured requirements documents. This LLM is augmented with knowledge, heuristics, and instructions that are pertinent to the software development process, requirements analysis, object-oriented design, and test-driven development, effectively emulating the expertise of a seasoned software engineer. We introduce a ``Progressive Prompting'' method that allows software engineers to engage with this LLM in a stepwise manner. Through this approach, the LLM incrementally tackles software development tasks by interpreting the provided requirements to extract functional requirements, using these to create object-oriented models, and subsequently generating unit tests and code based on the object-oriented designs. We demonstrate the LLM's proficiency in comprehending intricate user requirements and producing robust design and code solutions through a case study focused on the development of a web project. This study underscores the potential of integrating LLMs into the software development workflow to significantly enhance both efficiency and quality. The tailored LLM is available at \href{https://chat.openai.com/g/g-bahoiKzkB-software-engineer-gpt}{https://chat.openai.com/g/g-bahoiKzkB-software-engineer-gpt}.
\end{abstract}

\begin{IEEEkeywords}
Requirements Engineering, Large Language Models (LLMs), ChatGPT, Code Generation, Use Cases, Software Specification, Automated Software Engineering
\end{IEEEkeywords}

\section{Introduction}
In the realm of Software Engineering, Requirements Engineering (RE) is pivotal, serving as the foundation upon which all subsequent development stages are built. Requirements are traditionally documented in natural language \cite{kassab2014state}, varying from structured formats like use cases to informal descriptions. The textual nature of requirements inspires researchers to apply natural language processing technologies for various requirements engineering tasks \cite{zhao2021natural}. The advent of large language models (LLMs) such as ChatGPT offers a promising avenue for assisting requirements engineering processes. These language models, powered by deep learning algorithms and a gigantic training corpus, exhibit an unprecedented ability to understand, analyze, and generate natural language.

Despite the promising capabilities of LLMs, their application in translating natural language requirements into executable software code remains largely unexplored. This gap in application presents a unique opportunity to revolutionize the way software is developed. In this research preview, we present an initial attempt to use LLMs to assist software engineers by progressively refining use cases to functional requirements, design models, test cases, and eventually implementation code.

Our work is built on two assumptions. First, detailed requirements are provided to the LLMs as input. Second, software engineers' collaboration with LLMs is critical.

LLMs need detailed requirements as input. The recent hype in generative AI makes people have unrealistic expectations for LLMs. People envision that providing a brief description of a software project will enable LLMs to automatically generate a fully functional software system. However, LLMs are not mind readers. A critical aspect of leveraging LLMs is the quality of the input. In the context of software development, that input would be the requirements. The potential of the LLMs to generate effective code hinges on the details and clarity of the requirements documents. Well-crafted requirements are not just a necessity but a catalyst in fully harnessing the capabilities of LLMs. High-quality requirements provide a detailed and clear set of instructions. LLMs, with their advanced natural language processing capabilities, can interpret these instructions accurately, leading to more precise and effective code generation. This will make the software more aligned with the stakeholders' intent.

The second assumption is that LLMs need software engineers' collaboration to be effective. Instead of letting an LLM take full control, they should behave like a consultant for the software engineers. The LLMs and the software engineers ought to collaboratively find out the design details that will be used in the implementation of the software system. By automating parts of the analysis, design, and coding process, LLMs allow software engineers to focus on more complex and creative aspects of software development.

The contributions of this paper are as follows:
\begin{itemize}
  \item We introduced a tailored LLM for generating source code based on the provided software requirements. The LLM is fine-tuned on the knowledge, instructions, and heuristics pertinent to the software development process, requirements analysis, object-oriented design, and test-driven development. The LLM is available at \href{https://chat.openai.com/g/g-bahoiKzkB-software-engineer-gpt}{https://chat.openai.com/g/g-bahoiKzkB-software-engineer-gpt}.
  \item Our innovative integration of stepwise refinement of requirements into the LLM's design process significantly enhances its robustness and accuracy.
\end{itemize}

The structure of this paper is organized in the following manner: Section \ref{relatedWork} examines the existing literature on generating code with LLMs. Section \ref{researchMethod} presents a detailed description of our research method, outlining the systematic approach we have developed to enhance the application of LLMs in software engineering. Section \ref{tailoredLLM} elaborates on the design specifics of our custom-developed LLM. Following that, Section \ref{caseStudy} offers a case study that exemplifies the model's capability in comprehending requirements and producing code that is both syntactically and semantically accurate. Section \ref{discussion} explores the implications of LLMs within the field of software engineering and outlines the directions for future research. Finally, Section \ref{conclusion} concludes the paper.

\section{Related Work} \label{relatedWork}
Belzner and his colleagues presented an insightful analysis of the potential benefits and challenges of adopting LLMs in software engineering \cite{belzner2023large}. Their discussion spanned across areas such as requirements engineering, system design, code and test generation, code quality reviews, and software process management. The authors elucidated their position through a case study on the development of a ``search and rescue'' simulation for autonomous planning agents by interacting with a base ChatGPT model and a Google Bard model. They started with a one-line textual requirement, and after a series of conversations, the LLM generated simple test cases for the simulation. Hong and his colleagues \cite{hong2023metagpt} proposed an LLM-based multi-agent framework, MetaGPT, that considers the context of a software development process and assigns different roles (product manager, architect, engineer) to several agents. Their approach decomposed the complex task of developing software into manageable subtasks that require collaboration among agents. The open-source project, gpt-eningeer\footnote{\href{https://github.com/gpt-engineer-org/gpt-engineer}{https://github.com/gpt-engineer-org/gpt-engineer}}, allows a user to specify some high-level requirements like ``Create a multiplayer snake game in the browser using a Python back end with MVC components'', the LLM analyzes the requirements, asks for clarification, and then builds it.

The work referenced prior suggests a misconception that mere brief descriptions can compel an LLM to ``magically'' generate a fully operational software system. We challenge the notion of the so-called ``one-line requirement'' method in LLM-driven software development. The case studies cited to illustrate this approach often involve elementary and well-documented examples, such as the creation of a snake game. The simplicity of these examples and the widespread availability of their source codes online may mislead some to overestimate the efficacy of succinct requirement statements. While ``one-line requirement'' might suffice for tasks that are detailed and low-level, e.g., ``Create a web page to measure user reaction time by clicking a button''—which may directly leverage technical descriptions likely present in training datasets, their effectiveness diminishes with the complexity and uniqueness of software projects not represented in the LLM's training data. This paper aims to explore the boundaries of LLM capabilities in software development under such conditions.

A pivotal factor in employing LLMs for software development is the quality of input requirements. The model's success in accurately interpreting and translating these requirements significantly depends on their clarity and specificity. However, this critical aspect is often overlooked in existing literature. Our position is that human involvement in crafting requirements is irreplaceable by LLMs. While these models can assist in the process, their outputs must be meticulously reviewed and validated by human stakeholders. We advocate against the reliance on ``one-line requirement'' methodologies for LLM-based software development. Instead, we emphasize the importance of dedicating time and effort to articulate high-quality, detailed requirements and to thoroughly consider potential scenarios. Our version of LLM is optimized to comprehend structured and comprehensive inputs, rather than simplistic, one-liner prompts.

The only piece of related work that aligns with our perspective is the open-source initiative, GPT-Synthesizer\footnote{\href{https://github.com/RoboCoachTechnologies/GPT-Synthesizer}{https://github.com/RoboCoachTechnologies/GPT-Synthesizer}}. This tool stands out by assisting users in comprehending their project's needs and exploring design options through a well-organized interview process. Tailored for individuals unsure of how to initiate or define their software projects, the GPT Synthesizer proves to be a crucial aid.

\section{Research Method} \label{researchMethod}
The review of existing literature reveals a significant gap in the application of LLMs for the direct generation of executable software code from natural language requirements. To bridge this gap, our research introduces a novel method that not only enhances the LLM's understanding of software requirements but also iteratively refines its output to ensure that the generated code meets the semantic expectations. This methodological innovation is vital for advancing the use of LLMs in software development, particularly in environments where accuracy and alignment with business goals are paramount.

\subsection{Research Objectives}
Our research aims to develop and evaluate a tailored LLM that can effectively translate detailed software requirements into high-quality, functional code. The primary objective of this research is to demonstrate the feasibility of using LLMs for end-to-end software development tasks, thereby minimizing human intervention in the coding process while ensuring that the outputs are practically viable and aligned with user intentions.

\subsection{Research Design}
This research follows a phased approach:
\begin{itemize}
    \item Phase 1: Model Selection and Customization. We begin by selecting ChatGPT as the foundational LLM due to its advanced capabilities in understanding and generating natural language. The LLM is then customized to suit the specific needs of software development, integrating software engineering knowledge and instructions that enhance its ability to process and translate software requirements into functional code.
    \item Phase 2: Data Preparation. Software requirements will be collected with the intent to represent the diversity of real-world software development projects as accurately as possible. These will include both industry-standard specifications and hypothetical scenarios designed to test the model's capabilities across different scenarios.
    \item Phase 3: Evaluation and Refinement. The tailored LLM's effectiveness will be evaluated through a series of controlled tests, analyzing both the code quality and its alignment with the input requirements. The evaluation phase will involve comparing the LLM-generated code with expert-written code to assess accuracy, adherence to requirements, and practical usability. Both qualitative and quantitative measures will be employed to provide a thorough evaluation of the tailored LLM's performance. Feedback from these evaluations will be used to further refine the model.
\end{itemize}

\subsection{Iterative Refinement Process}
An iterative refinement process is integral to the methodology. After each evaluation phase, detailed feedback will be analyzed to identify any shortcomings of the tailored LLM. This feedback informs subsequent refinements to the model, enhancing the tailored LLM's ability to generate more accurate and relevant code outputs.

\section{A Tailored LLM for Software Engineering} \label{tailoredLLM}
\begin{figure}[htbp]
\centerline{\includegraphics[width=\linewidth]{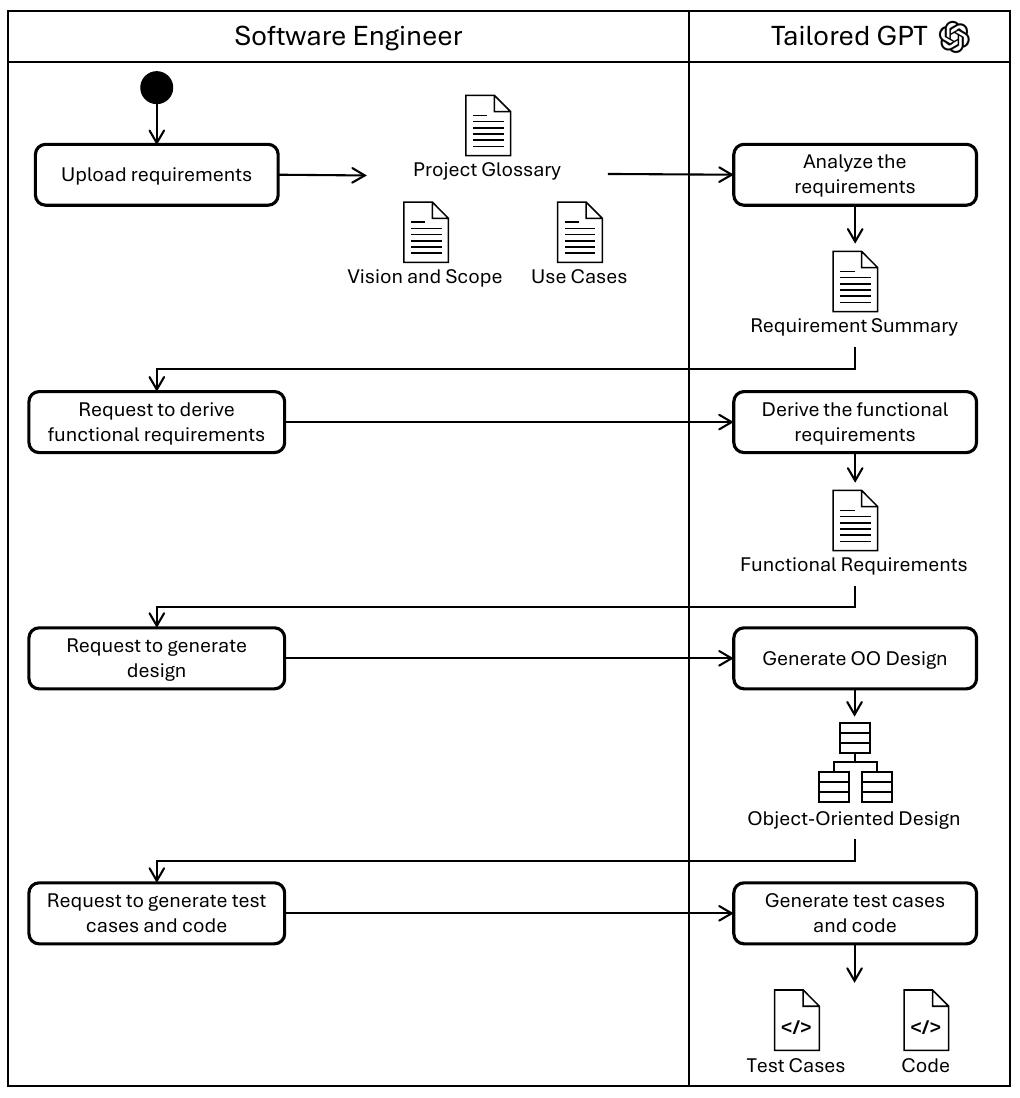}}
\caption{A human software engineer's interaction with the tailored GPT model.}
\label{figure:workflow}
\end{figure}

In this section, we explore the design of our tailored LLM crafted to support software engineers in generating code from high-level requirements. To depict the process of engagement with our customized LLM, we employ a UML activity diagram, which outlines the workflow for interaction between the software engineers and the LLM (Fig. \ref{figure:workflow}).

\subsection{Software Requirements Documents}
Without a foundation of high-quality requirements as input, even the most performant LLM can be rendered useless. Instead of typing a few informal itemized requirements to an LLM, we require the software development team to upload to the LLM structured requirements in the format of project glossary (\href{https://docs.google.com/document/d/1yqUkax6duHvEfFCP50IP16yoxQGyDaMHBT-CxUA73zk/edit?usp=sharing}{Google Docs}), vision and scope (\href{https://docs.google.com/document/d/1h7Bho4auvUAE5zuPNh4Jkk0TBSNfom8P8F_e11kugqY/edit?usp=sharing}{Google Docs}), and use cases (\href{https://docs.google.com/document/d/1vaoprKQn58N4uE5gLaqLFLWtvsBYralnNWP_7rj4vJY/edit?usp=sharing}{Google Docs}).

\begin{itemize}
    \item Project Glossary: This document contains definitions of all key terms used within the project, ensuring clarity and consistency across all project communications and artifacts.
    \item Vision and Scope: This outlines the overall objectives, the scope of the project, and the roles of different stakeholders, providing a comprehensive overview of the project's goals and expected outcomes.
    \item Use Cases: This document includes detailed descriptions of each use case, illustrating specific functionality that the software is expected to perform. These are particularly valuable for understanding the interactions between the system and its users, as well as between different components of the system.
\end{itemize}

The time and effort invested in writing high-quality software requirements pays off significantly, especially in the era of LLMs.

\subsection{A Progressive Prompting Approach}
Inspired by the stepwise refinement technique in program development, we adopt a progressive prompting approach to interact with the LLMs. In Fig. \ref{figure:workflow}, the interaction starts with the software engineer uploading detailed requirements documents to the LLM. Instead of instructing the LLM to rush to the final artifact i.e., the test cases and code, we give the LLM ``time to think'' by asking the LLM to follow a waterfall software development process and come up with intermediate artifacts along the way. In particular, the software engineer prompts the LLM to first refine requirements into detailed functional requirements. Then, the LLM is prompted to devise an object-oriented design based on the functional requirements. The purpose of the design is to discover objects, assign responsibilities to them, and model their interactions. Finally, based on the functional requirements and the design produced in the previous steps, the LLM is prompted to generate test cases and code.

\begin{figure}[htbp]
\centerline{\includegraphics[width=0.6\linewidth]{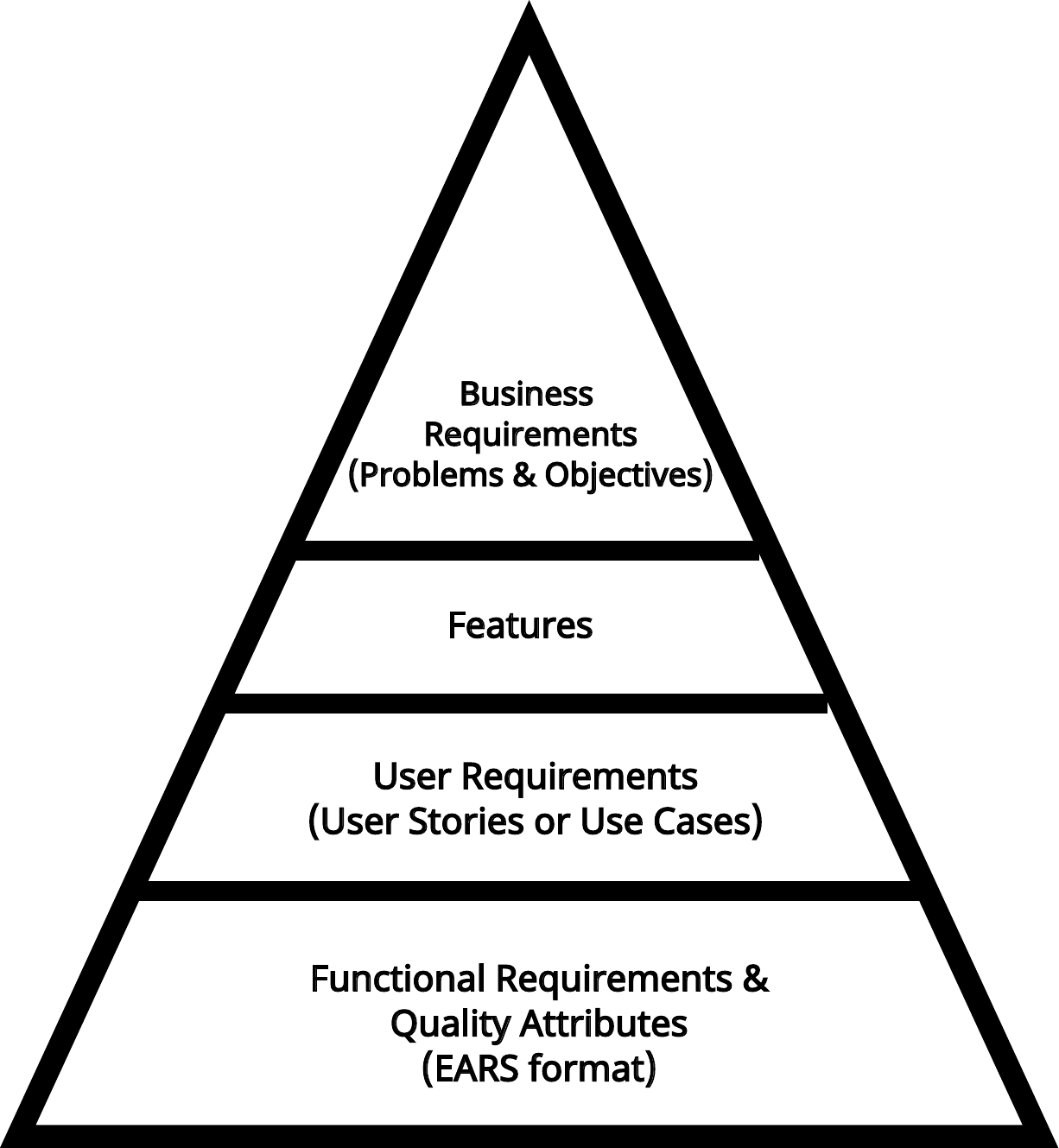}}
\caption{A requirement hierarchy.}
\label{figure:hierarchy}
\end{figure}

This progressive prompting approach is in line with both the waterfall software development process and the requirements abstraction hierarchy (Fig. \ref{figure:hierarchy}). By breaking down the complex software development task into smaller and more specific subtasks, this approach generates a ``chain of thought'' \cite{wei2022chain}, making the LLM more efficient and effective in performing complex tasks in software development.
The intermediate artifacts like functional requirements and object-oriented design are valuable to the software development team since they allow the human engineers to understand and verify the LLM's chain of thought and provide feedback. Those artifacts also help verify in case of hallucination. This will also have the benefit of requirement traceability \cite{pinheiro2004requirements}.

\subsection{Instructions and Knowledge Base for the Tailored LLM}
In this study, we selected ChatGPT as the foundational LLM. Prompts can be utilized in two ways when interacting with an LLM like ChatGPT. The first method involves directly requesting ChatGPT to generate code based on provided requirements. The second method introduces a structured approach where, instead of allowing ChatGPT to autonomously navigate software development tasks, we embed specific instructions to guide the development process. These instructions include, for example, heuristics on converting use cases into functional requirements and developing an object-oriented design. These guidelines act as constraints and safeguards to align the development closely with an organization's standard processes and mitigate the risk of generating irrelevant or inaccurate content.

In this work, we took advantage of ChatGPT's new feature, GPTs, by customizing a base ChatGPT to excel in generating code from high-level requirements. The tailored GPT is named ``Software Engineer GPT''\footnote{\href{https://chat.openai.com/g/g-bahoiKzkB-software-engineer-gpt}{https://chat.openai.com/g/g-bahoiKzkB-software-engineer-gpt}}. We also make the instructions for the GPT available\footnote{\href{https://github.com/Washingtonwei/software-engineer-gpt}{https://github.com/Washingtonwei/software-engineer-gpt}} so that other practitioners can freely modify the instructions to create their own customized GPT.

Besides instructions, we provided a comprehensive knowledge base\footnote{\href{https://github.com/Washingtonwei/software-engineer-gpt/blob/main/knowledge.md}{https://github.com/Washingtonwei/software-engineer-gpt/blob/main/knowledge.md}} for the tailored GPT. The document is an extensive exploration of requirements engineering. It starts with a general introduction to requirements engineering, emphasizing its role as a collection of activities for defining, documenting, and adapting the objectives and constraints of a software system based on existing problems and new technological opportunities. The document then delves into the specific roles and responsibilities of a business analyst in this process, detailing their crucial function in forging partnerships between customers and developers. It covers various types of requirements in software projects, such as business requirements, user requirements, functional requirements, and quality attributes. These are organized into a hierarchy, illustrating how each level adds detail and elaborates on higher-level requirements. The document also provides an in-depth look at user requirements, including their representation through use cases, and the specific elements that constitute a use case. Additionally, it addresses functional requirements and their derivation from user requirements. Furthermore, the document outlines a software development workflow incorporating requirements analysis, object-oriented design, and testing, followed by a discussion on software architecture with a focus on the ``Clean Architecture''. Finally, the document offers guidelines for object-oriented design, including identifying key classes and assigning responsibilities, and concludes with an overview of test-driven development, demonstrating its application in the context of the presented software development approach. The contents in the knowledge base are extracted from \cite{wiegers2013software} and \cite{meyer1997object}.

\section{Case Study} \label{caseStudy}
To demonstrate LLMs' capability of generating code from requirements, we chose a medium-sized web project called ``SuperFrog Scheduler''. SuperFrog is the iconic horned frog mascot of Texas Christian University (TCU), known for its participation in numerous events annually, including sports, weddings, charity functions, and media appearances. To facilitate the scheduling of SuperFrog's appearances, a request management system is necessary, involving submission by customers, review by the Spirit Director, and acceptance by student performers embodying SuperFrog. A development team, consisting of five software developers, commenced work on the SuperFrog Scheduler in October 2018 and completed the project in May 2019. During the requirements gathering phase, the team identified and documented 26 user-goal level use cases, which are available here (\href{https://docs.google.com/document/d/1PBDgqCbMPpyrAWZnob_OxDecOPoyyOtV_sJeeObh62E/edit?usp=sharing}{Google Docs}). The ``SuperFrog Scheduler'' was selected for its complexity and relevance, involving multiple stakeholders and intricate scheduling requirements typical of real-world software projects. This case study not only provides detailed documentation for verifying the tailored LLM's code generation capabilities but also showcases the model's practical application in a dynamic educational environment. Its visibility and the educational context offer a unique opportunity to demonstrate the potential of LLMs in streamlining complex software development tasks while educating future software engineers on AI-driven development tools. For this preview paper, our focus is restricted to the implementation of Use Case 18 (\href{https://docs.google.com/document/d/1PBDgqCbMPpyrAWZnob_OxDecOPoyyOtV_sJeeObh62E/edit#bookmark=id.ff9ddug0ipbh}{Google Docs}). Use Case 18 describes the process where the Spirit Director generates TCU Honorarium Payment Request Forms for SuperFrog Students who have completed appearance requests. The system facilitates the creation of these forms by providing the necessary details for each eligible appearance. This process aims to ensure SuperFrog Student workers are compensated for their services in a timely and efficient manner.

Note that the design and the source code of this project are never made accessible via the web, and therefore not part of ChatGPT's training data. This is different from all the previous work. In this section, we demonstrate the complete software development process focusing on Use Case 18 by conversing with the tailored GPT. The complete conversation with the tailored GPT is available at this \href{https://chat.openai.com/share/8f52df6a-d00c-44df-a2c4-4a9235415418}{link}\footnote{\href{https://chat.openai.com/share/8f52df6a-d00c-44df-a2c4-4a9235415418}{https://chat.openai.com/share/8f52df6a-d00c-44df-a2c4-4a9235415418}}. This real-world application can provide tangible evidence of the effectiveness and challenges of using ChatGPT in software engineering.

To enhance the accessibility and reproducibility of our research, we have compiled all essential project artifacts into a replication package\footnote{\href{https://drive.google.com/drive/folders/1CFxMNblwOV-mthVcRRb2gTsFxNRO7nlz?usp=sharing}{https://drive.google.com/drive/folders/1CFxMNblwOV-mthVcRRb2gTsFxNRO7nlz?usp=sharing}}. This package includes the project glossary, vision and scope, and detailed use cases, which are fundamental to understanding the development context and requirements of the ``SuperFrog Scheduler''.

\subsection{Uploading Software Requirements}
\begin{figure}[htbp]
\centerline{\includegraphics[width=\linewidth]{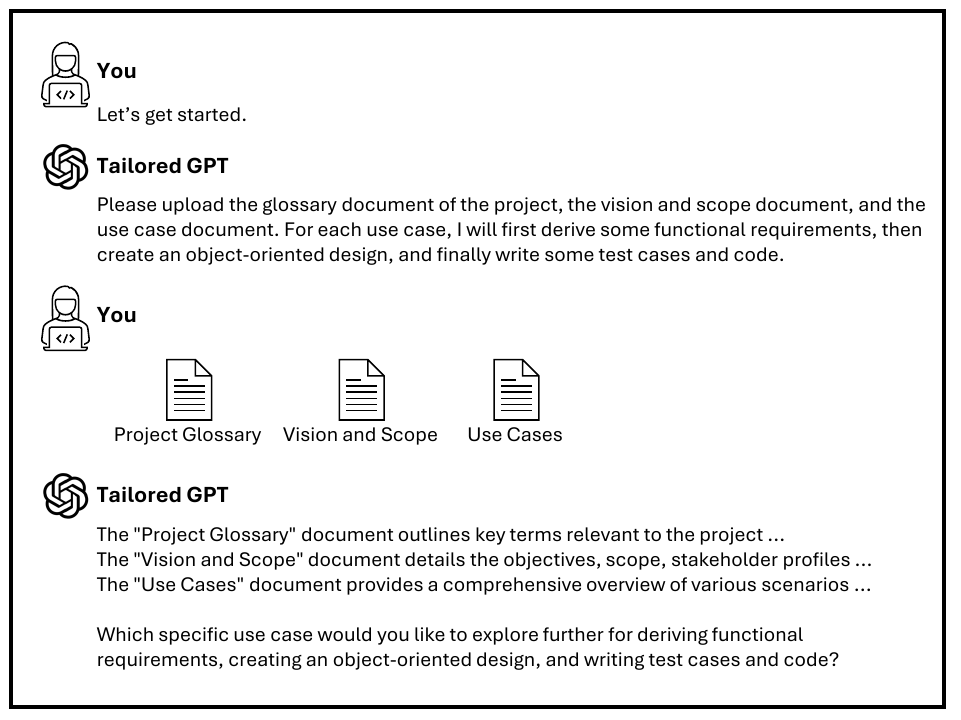}}
\caption{The user uploads requirements documents to Software Engineer GPT.}
\label{figure:upload}
\end{figure}

We initiated the process by uploading the project glossary, vision and scope, and use cases to the tailored GPT. These requirements documents provide critical context and background information for subsequent tasks needed. The GPT then summarized the contents of the three documents (see the conversation excerpt in Fig. \ref{figure:upload}). The GPT prompted us to pick one use case to start.

\subsection{Deriving Functional Requirements}
\begin{figure}[htbp]
\centerline{\includegraphics[width=\linewidth]{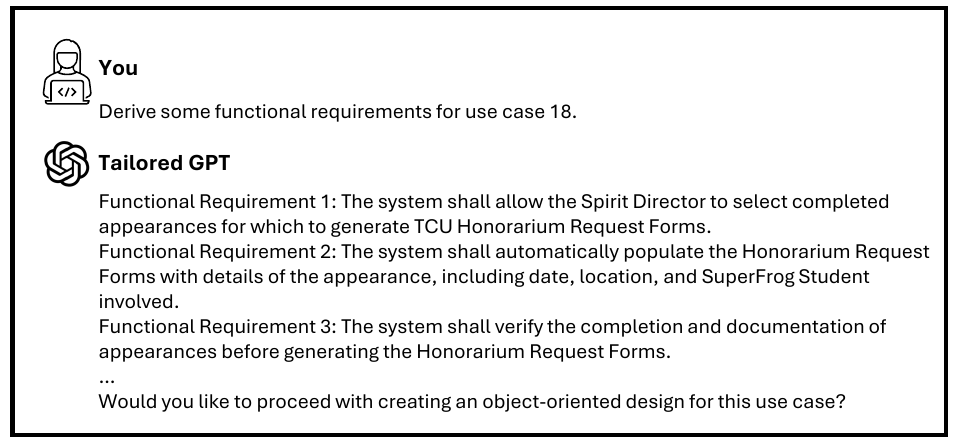}}
\caption{Software Engineer GPT derives the functional requirements from use cases.}
\label{figure:deriveFunctionalRequirements}
\end{figure}

Use cases describe tasks that users will need to perform with the system, or user-system interactions that will result in a valuable outcome for some stakeholders. We instructed the GPT to derive the necessary functional requirements that must be implemented to enable each use case. As a proof of concept, we only demonstrated the design of \href{https://docs.google.com/document/d/1PBDgqCbMPpyrAWZnob_OxDecOPoyyOtV_sJeeObh62E/edit#bookmark=id.ff9ddug0ipbh}{Use Case 18}, which pertains to the generation of Honorarium (Payment for Services) request forms for student workers (see the derived functional requirements in the conversation excerpt in Fig. \ref{figure:deriveFunctionalRequirements}). Then, the GPT asked us to proceed with creating an object-oriented design for this use case.

\subsection{Generating an Object-Oriented Design}
In this step, we instructed the GPT to carry out an object-oriented design for Functional Requirement 2: The system shall automatically populate the Honorarium Request Forms with details of the appearance, including date, location, and SuperFrog Student involved. Based on the instructions defined in the knowledge document, the GPT:
\begin{itemize}
  \item identified key classes that users use to describe the problem domain and implementers use to describe the solution domain,
  \item assigned responsibilities to each class,
  \item provided properties and operations to each class that are needed to carry out these responsibilities,
  \item came up with an object collaboration so that the objects of the classes can collaborate to carry out the functional requirements of a use case.
\end{itemize}

\begin{figure}[htbp]
\centerline{\includegraphics[width=\linewidth]{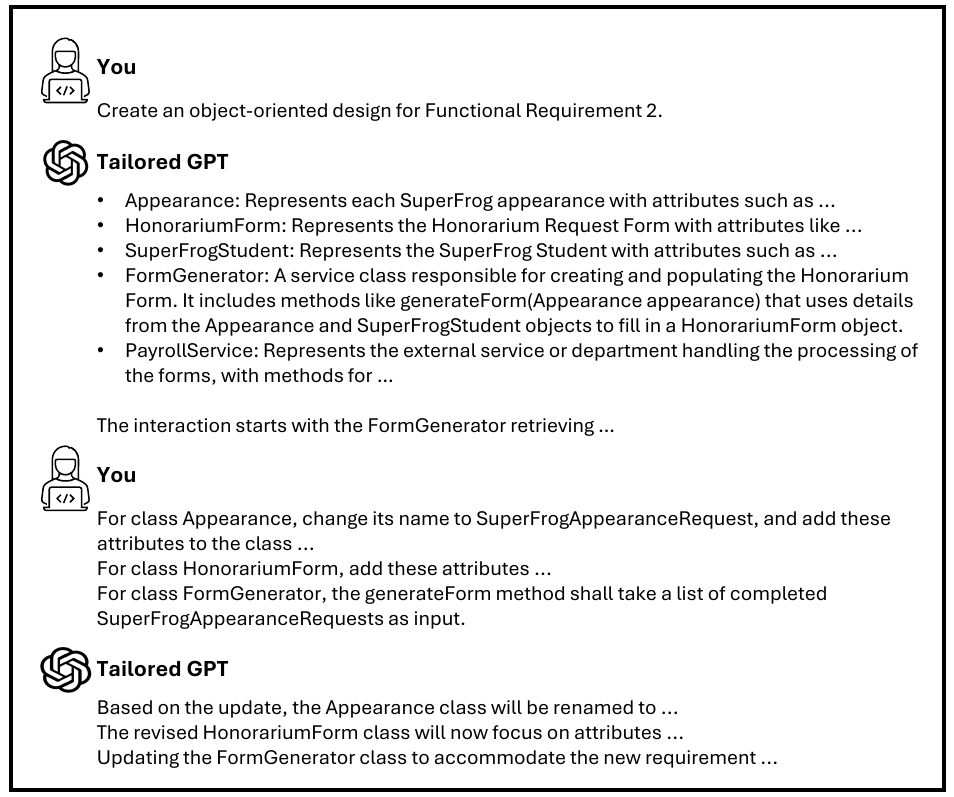}}
\caption{Software Engineer GPT creates an object-oriented design for one functional requirement.}
\label{figure:ood}
\end{figure}

Fig. \ref{figure:ood} shows the excerpt of the conversation. As expected, some designs were not aligned with our intentions. So, we provided detailed feedback to the GPT, and the GPT quickly adjusted the design.

\subsection{Generating Test Cases}
\begin{figure}[htbp]
\centerline{\includegraphics[width=\linewidth]{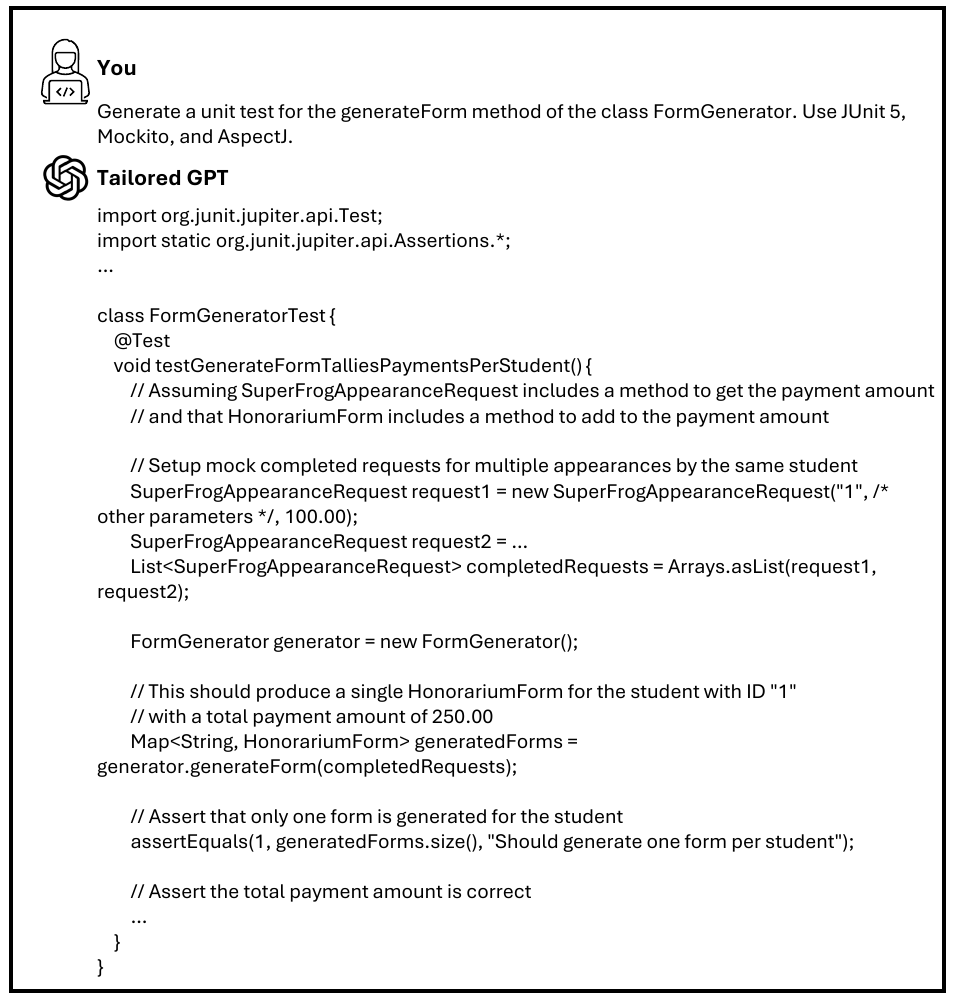}}
\caption{Software Engineer GPT generates a test case for one method of a class.}
\label{figure:tests}
\end{figure}

In this step (Fig. \ref{figure:tests}), the tailored GPT was instructed to generate one unit test case for the generateForm method in the class FormGenerator. We also specified the technologies required. It turned out that again we needed to clarify some misunderstandings (detailed conversion is not shown).

\subsection{Generating Implementation Code}
\begin{figure}[htbp]
\centerline{\includegraphics[width=\linewidth]{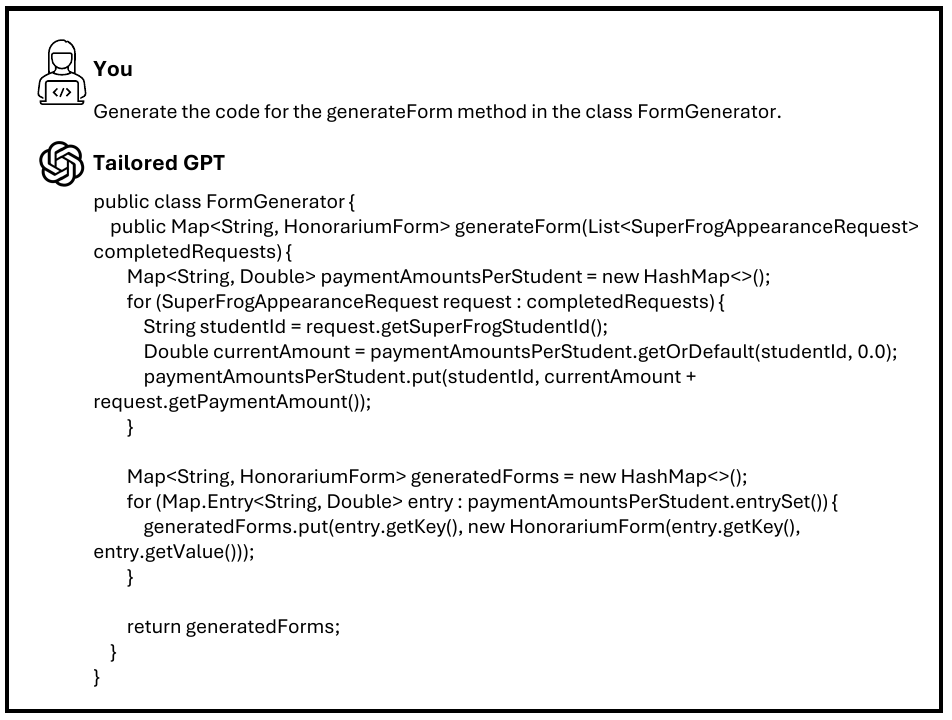}}
\caption{Software Engineer GPT generates the implementation code based on the test case.}
\label{figure:code}
\end{figure}

In the end, based on the intermediate artifacts or ``chain of thoughts'' generated in the previous steps, we believe that the GPT has enough context and information to move to the final step. We then instructed the GPT to generate the implementation code for the generateForm in the class FormGenerator (Fig. \ref{figure:code}).

\section{Discussion} \label{discussion}
In this section, we explore the implications of our approach for the future of software engineering, education, and agile methodology. Additionally, we address the limitations of our study and outline directions for future research.

\subsection{Requirements are All You Need}

While the title of this paper, ``Requirements are All You Need: From Requirements to Code with LLMs'', might seem hyperbolic, it aptly highlights the pivotal role of requirements engineering in contemporary AI-assisted software engineering. Even amidst the advanced capabilities of LLMs and other AI tools in streamlining the coding process, the initial step of meticulously defining requirements remains a cornerstone. A well-articulated set of requirements is fundamental to leveraging AI technologies effectively in software development, thereby reinforcing the indispensability of requirements engineering as a discipline within the AI-enhanced software engineering landscape.

\subsection{Impact on Software Engineering Education}
LLMs have transformed the landscape of programming, showcasing exceptional abilities in code generation and problem-solving. This technological advancement has undeniably increased efficiency and productivity within software development. However, the question arises: Is teaching traditional designing, programming, and testing skills to students still necessary? The unequivocal answer is yes. Despite the prowess of LLMs in handling a variety of programming tasks, their limitations cannot be overlooked. These models sometimes misunderstand us and produce code that is not aligned with our intent, underscoring the need for human oversight and validation. It is imperative for software engineering education to evolve, teaching students to critically assess AI-generated designs and code for accuracy, efficiency, maintainability, and security.

The emergence of LLMs, such as ChatGPT, and tools like GitHub Copilot and Devin AI, signals a pivotal shift in software engineering education. To stay ahead, educators must not only embrace the LLMs themselves but also guide students on how to effectively leverage LLMs. This entails a significant pedagogical shift towards enhancing soft skills. Students should be trained to excel in communication with various stakeholders, the art of drafting high-quality software requirements, and conducting thorough design and code reviews. Such skills are becoming increasingly vital as the role of manual coding, particularly for minor features, diminishes.

\subsection{LLMs Redefine Agile}
This paper presupposes the availability of detailed and extensive requirements, such as use cases, for input into LLMs. Crafting effective requirements, however, can be a labor-intensive process, and this strategy might initially be met with skepticism. A key question arises: How does this strategy align with Agile methodologies, which prioritize working software over comprehensive documentation? Critics may argue that the focus on detailed and often extensive use cases could result in an undue emphasis on documentation, thereby potentially redirecting time and resources away from actual software development. Yet, this concern may be mitigated in the context of LLMs. We contend that the emergence of LLMs is poised to transform the Agile methodology. Owing to LLMs' capabilities in test and code generation, the process of translating well-defined requirements into code has become significantly more efficient. This efficiency could allow software engineers to allocate more time to the crucial tasks of communication and requirement specification, thereby enhancing the development process.

\subsection{Limitations and Future Work}
The proof-of-concept case study showcased the significant potential of leveraging LLMs in software development, particularly in the context of use-case-driven development for web projects. However, further testing across a broader spectrum of software projects—including real-time systems, AI-driven applications, and games—is essential to fully understand the applicability and limitations of LLMs in these varied contexts. Additionally, exploring different requirements specifications and design methodologies is crucial to assess the adaptability and effectiveness of LLMs in diverse software engineering paradigms.
While a tailored version of ChatGPT has been developed and made publicly available, utilizing the default user interface of ChatGPT for software development presents practical challenges. Specifically, software engineers may find it cumbersome to navigate through previously generated designs or code snippets due to the need to scroll extensively. To address this issue, we envisage the creation of a user-friendly web application in future work. This application will aim to streamline the process, offering a more intuitive and visual approach to translating requirements into tests and code.

A comprehensive evaluation of this approach remains a priority for future research. We intend to encourage software engineers to incorporate these LLMs into their workflow and solicit their feedback. This will provide valuable insights into the practicality, efficiency, and user experience of using LLMs in real-world software development scenarios, thereby informing further improvements and adaptations.

\section{Conclusion} \label{conclusion}
In this research preview, we introduced a tailored LLM designed to assist software engineers in refining requirements into software design, test cases, and implementation code. This LLM incorporates an external knowledge base that encompasses both the knowledge and heuristics of requirements engineering and software design. A case study was conducted to demonstrate the practicality and effectiveness of this customized LLM, underscoring its potential in software development. Furthermore, our research underscored the critical role of meticulously crafted requirements in maximizing the benefits of LLMs within the realm of software engineering.

Future work will explore a broader array of requirements specification techniques and software design methodologies. Additionally, a more thorough evaluation of this innovative approach will be undertaken to further validate its efficacy and applicability in diverse software engineering contexts.

\bibliographystyle{IEEEtran}
\bibliography{IEEEabrv,IEEEexample}

\vspace{12pt}

\end{document}